\documentclass[lengthcheck,onecolumn,prfluids,superscriptaddress,showpacs,floatfix,aps,longbibliography]{revtex4-2}
\usepackage{amsmath,amssymb,amsfonts,latexsym,bm}
\usepackage{graphicx,epsfig}
\usepackage{fancyvrb}
\usepackage{booktabs}
\usepackage{soul}
\usepackage{ragged2e}
\usepackage[svgnames]{xcolor}
\usepackage{xfrac}
\usepackage{mathrsfs}
\usepackage{url}
\usepackage[colorlinks,linkcolor=MediumBlue,citecolor=MediumBlue,urlcolor = MediumBlue]{hyperref}

\allowdisplaybreaks

\begin{document}

\title{Acceleration of Lagrangian Particles in Shell Models of Turbulence}

\author{Lorenzo Piro}\email{lorenzo.piro@roma2.infn.it}
\affiliation{Department of Physics \& INFN, University of Rome ``Tor Vergata", Via della Ricerca Scientifica 1, 00133 Rome, Italy}
\author{Massimo Cencini} 
\affiliation{Istituto dei Sistemi Complessi, CNR, Via dei Taurini 19, 00185 Rome, Italy} \affiliation{INFN ``Tor Vergata", Via della Ricerca Scientifica 1, 00133 Rome, Italy}
\author{Roberto Benzi} 
\affiliation{Sino-Europe Complex Science Center, School of Mathematics \\North University of China, Shanxi, Taiyuan 030051, China} 
\affiliation{Department of Physics \& INFN, University of Rome ``Tor Vergata", Via della Ricerca Scientifica 1, 00133 Rome, Italy}

\date{\today}

\begin{abstract}
Lagrangian acceleration has been investigated both experimentally and numerically in the past, and it has been shown to exhibit extreme fluctuations, which have been rationalized as events in which tracer particles get trapped into vortical structures such as vortex tubes or filaments. Here, we consider the statistics of acceleration within the multifractal framework, as in Biferale et al. \textit{Phys. Rev. Lett.} \textbf{93} 064502 (2004), and investigate the statistics of Lagrangian acceleration using shell models of turbulence, as in G. Boffetta et al. \textit{Phys. Rev. E} 66, 066307 (2002), that -- by construction -- do not contain vortical structures. Our analysis reveals that, despite not accounting for specific coherent vortex structures, the multifractal model accurately captures the extreme intermittent fluctuations observed in the acceleration, with predictions that remain robust across a wide range of Reynolds numbers.
\end{abstract}
 
\maketitle

\textit{Introduction.} 
In the last few decades, many properties of fully developed three-dimensional turbulence have been disentangled. Limiting ourselves to the case of homogeneous and isotropic turbulence (HIT), the original picture proposed by the celebrated Kolmogorov theory (K41)~\cite{Kolmogorov41} has been thoroughly revised to properly account for intermittency~\cite{Frisch1995}. Indeed, while K41 theory predicts that turbulence should exhibit scale-invariant properties, laboratory experiments and high-resolution direct numerical simulations (DNS) of the Navier-Stokes equations show substantial violations of scale invariance in the statistical properties of the velocity field~\cite{arneodo1996} and the local rate of energy dissipation~\cite{anselmet2001turbulent} referred to as {\it intermittency}. For passive scalar advected by turbulent velocity field -- and, in particular, for
random velocity fields as for the so-called Kraichnan model~\cite{kraichnan1968small} -- it has been shown that scale invariance does not hold, and anomalous scaling of correlation functions has been predicted~\cite{gawedzki1995anomalous,chertkov1996anomalous,shraiman1996symmetry} and successfully compared with simulations~\cite{frisch1999lagrangian}. 
Notably, in passive scalar turbulence, anomalous scaling exponents have been theoretically traced back to inertial range dynamics, and they are universal, i.e., independent of large-scale forcing and small-scale dissipation mechanisms (see the reviews~\cite{shraiman2000scalar,falkovich2001particles}). 
The same results are believed to hold for Navier-Stokes turbulence, namely that the dynamics of the inertial range dictate the intermittent fluctuations, which are universal in the manner defined above. 

The multifractal theory of turbulence, originally proposed by Parisi and Frisch~\cite{Parisi85}, constitutes a very successful framework for rationalizing intermittency and anomalous scaling in the inertial range of turbulence. 
The multifractal theory exploits the invariance of the Navier-Stokes equations under the scaling transformation $r \rightarrow \lambda r$, $v \rightarrow \lambda^h v$, $t \rightarrow \lambda^{1-h} t$ supplemented by the probability $\sim r^{3-D(h)}$ for the exponent $h$ (see below for details). The function $D(h)$ is related to the intermittent fluctuations of energy transfer in the inertial range. Notably, the same picture applies to both Eulerian and Lagrangian quantities~\cite{mordant2001measurement}, and one can successfully predict anomalous exponents in the Lagrangian dynamics starting from the knowledge of anomalous scaling in the Eulerian frame~\cite{boffetta2002,arneodo2008}.

In the Lagrangian dynamics, strong intermittency is observed in the statistical properties of the acceleration, $a_L$, experienced by tracer particles~\cite{la2001fluid,mordant2004three,biferale2004,buaria2022}. Numerical simulations and laboratory experiments have shown that extreme intermittent events of $a_L$ occur whenever the particle experiences the effect of vortex filaments or vortex
tubes~\cite{la2001fluid,mordant2004three,biferale2005particle}. Actually, the dynamical time scale, $\tau_L$, of the Lagrangian acceleration is close to the Kolmogorov time scale, $\tau_K$, which characterizes the dynamics of the dissipative scales. Thus, in principle, Lagrangian acceleration is not an inertial range quantity. However, within the multifractal theory, the Kolmogorov scale $\eta_K$ and time $\tau_K$ (and thus $\tau_L$) are fluctuating quantities whose statistical properties are controlled by $D(h)$~\cite{FV1991,biferale1999exit,boffetta2008twenty}, and thus inherited from the inertial range physics. It turns out that the multifractal theory provides a prediction for the probability distribution $P(a_L)$, which depends on the Reynolds number $Re$ and the function $D(h)$ without any reference to the peculiar \textit{coherent structures} of the turbulent flows~\cite{biferale2004}. 

Then, a simple yet nontrivial question is whether or not extreme intermittent fluctuations of Lagrangian acceleration $a_L$ can be observed without the effect of vortex filaments or any other geometrical structure in the flow, provided the presence of anomalous scaling of the structure functions in the inertial range. This Letter aims to answer such a question. Clearly, this cannot be addressed using the Navier-Stokes equations since vortex filaments are always present in HIT. 
Thus, here, we consider shell model of turbulence for modeling also single-particle Lagrangian dynamics~\cite{boffetta2002}, where \textit{by construction} no well-defined structures such as vortex filaments do occur.

\textit{Shell model for Lagrangian dynamics.} 
Shell models~\cite{biferale2003,bohr2005} are dynamical systems constructed to mimic the basic phenomenology of the turbulent energy cascade on a discrete set of scales (or Fourier modes), $\ell_n=k_n^{-1}=2^{-n}$ ($n=0,\ldots, N$), such that for each scale $\ell_n$ the velocity fluctuation is represented by a single complex variable $u_n$. The geometrical progression of scales and the limited number of variables allow shell models to reach high Reynolds numbers currently unattainable in DNS. In particular, we consider the Sabra model~\cite{Itamar}, where the complex velocities evolve according to the dynamics
\begin{equation}
    \dot{u}_n=ik_n(u_{n+2}u^*_{n+1}-\frac{1}{4}u_{n+1}u^*_{n-1}+\frac{1}{8}u_{n-1}u_{n-2})-\nu k_n^2 u_n+f_n\,,
    \label{eq:Sabra}
\end{equation}
where the $^*$ denotes complex conjugation. Equation~\eqref{eq:Sabra} resembles the Navier-Stokes equation in Fourier space with the nonlinear term restricted to locally interacting shells. 
The forcing term $f_n$, acting at large scales, injects energy at a rate $\epsilon=\langle \sum_n \Re\{f_n u_n^*\}\rangle$, while the viscous term ($-\nu k_n^2u_n$) removes it at small scales. In this Letter, we considered two different kinds of forcings: a random Gaussian acting on the first shell, obtained by evolving a Langevin equation so as to have a finite correlation time, and a forcing with constant power input, acting on the first two shells. Numerically, we found the same quantitative results for both Eulerian and Lagrangian intermittency.

\bigskip

\begin{figure}[t!]
\centering
\includegraphics[width=\columnwidth]{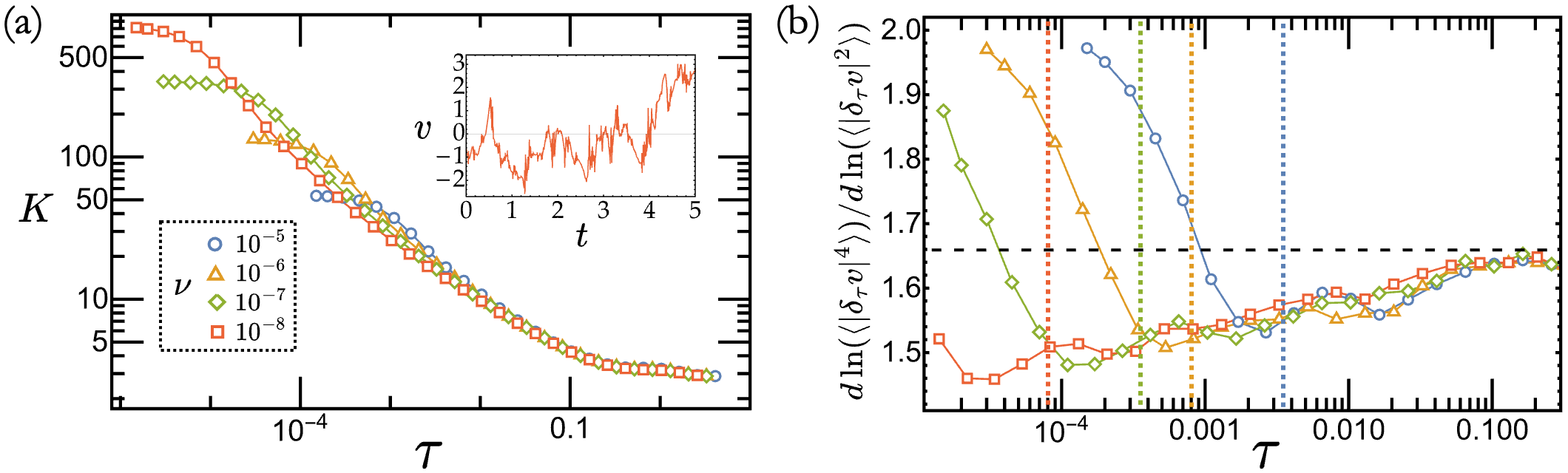}
\caption{\justifying (a) Kurtosis $K$ of the Lagrangian velocity increments as a function of the time lag $\tau$ obtained in the shell model for four values of the viscosity $\nu$: $10^{-5}$ (blue circles), $10^{-6}$ (orange triangles), $10^{-7}$ (green diamonds), and $10^{-8}$ (red squares). Inset: example of a typical time evolution of the Lagrangian velocity obtained from our numerical simulations with $\nu=10^{-8}$. (b) Local slopes as a function of the time lag $\tau$ for the same values of viscosity shown in (a). The horizontal dashed line represents the scaling behavior predicted by the multifractal theory.
The dotted vertical lines denote the value of $\tau_K=\sqrt{\nu/\epsilon}$.}
\label{fig:1}
\end{figure}

\begin{figure*}[t!]
\centering
\includegraphics[width=\textwidth]{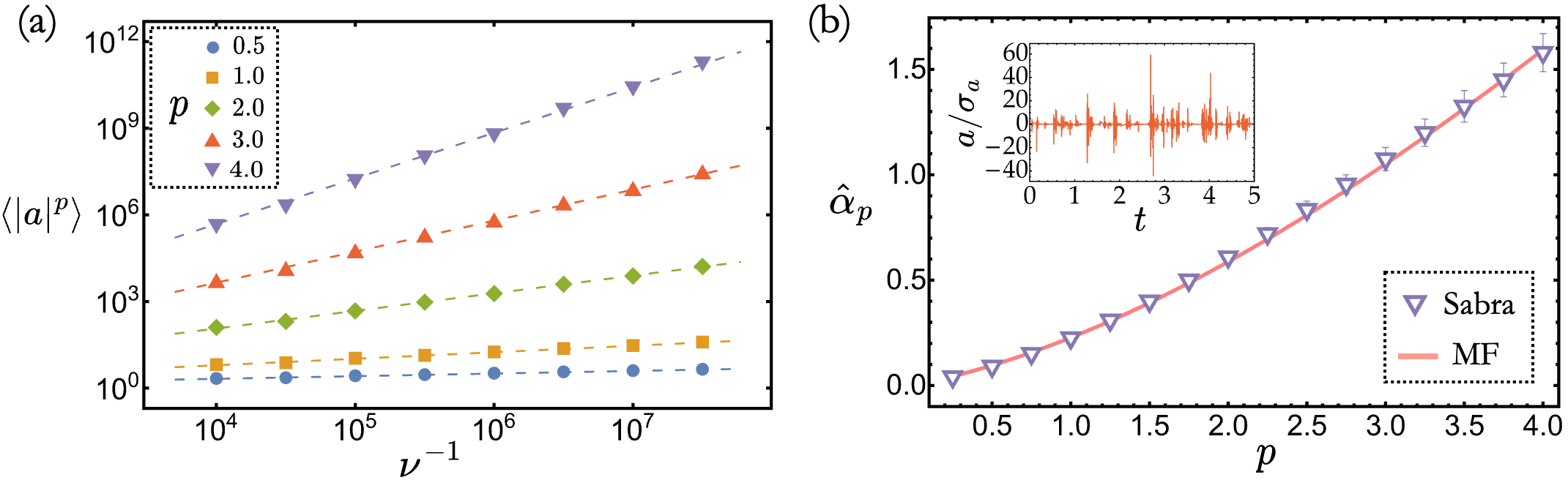}
\caption{\justifying (a) Moment of the acceleration, $\langle|a|^p \rangle$ defined in \eqref{eq:lagshell} vs $\nu^{-1}$ for different $p$, the dashed line represent the best fits. Notice that when varying $\nu$, we keep unchanged the large scales so that $Re\sim\nu^{-1}$ up to a multiplicative constant.
(b) Symbols show the fitted exponents $\hat{\alpha}_p$ (triangles) ruling the behavior of acceleration moments, $\langle|a|^p \rangle \sim Re^{\hat{\alpha}_p}$, obtained in the shell model together with the multifractal prediction obtained using Eq.~\eqref{expacc} with the log-Poisson model \eqref{eq:SL} (solid curve). Inset: example of the typical time evolution of the Lagrangian acceleration normalized by its standard deviation obtained from our numerical simulations with $\nu=10^{-8}$. In both panels, error bars are within the symbols if not visible.}
\label{fig:2}
\end{figure*}

Shell models were originally introduced to mimic the Eulerian properties of turbulent flows. Following~\cite{boffetta2002}, we employ them to model the Lagrangian velocity and acceleration along a fluid particle as the sum of the real part of velocity and acceleration fluctuations, respectively, at all shells, i.e.,
\begin{equation}
    v(t)\equiv \sum_{n=1}^{N} \Re\{u_n\}\,, \quad a(t)\equiv \dot{v}(t)\equiv \sum_{n=1}^{N} \Re\{\dot{u}_n\}\,,
    \label{eq:lagshell}
\end{equation}
the inset of Fig.~\ref{fig:1}(a) and Fig.~\ref{fig:2}(b) show an example of the time evolution of $v$ and $a$, respectively, obtained with above summation. 
Indeed, the Lagrangian velocity can be represented as the superposition of velocity fluctuations generated by turbulent eddies, which also move with the advected tracers. As originally shown in~\cite{boffetta2002}, one may obtain more general representations by multiplying the shell‐model variables by suitably chosen wavelet functions. However, in the shell model, this superposition is most simply realized by the summation in Eq.~\eqref{eq:lagshell}. At any rate, $v(t)$ as defined by Eq.~\eqref{eq:lagshell} exhibits multifractal scaling in time, as first shown in~\cite{boffetta2002}.
Remarkably, both the Eulerian and Lagrangian velocity statistics of shell models are well captured by the multifractal model~\cite{boffetta2002}, whose basic ideas are summarized in the following. 

\textit{Multifractal model for Eulerian velocity statistics.} 
A straightforward way to detect intermittency in HIT is to consider the statistical properties of the longitudinal velocity difference $\delta_r v= [\vec v (\vec x + \vec r) - \vec v(\vec x)] \cdot \vec r / |\vec r |$.  The moments of $\delta_r v$ define the structure functions $S_p(r) \equiv \langle \delta_r v^p \rangle$.  There is strong numerical and experimental evidence~\cite{arneodo1996} (see also 
the review \cite{benzitoschi}) that in HIT, at large $Re$, $S_p(r)$ exhibit anomalous scaling, i.e.,
\begin{equation}
    S_p(r) \sim r^{\zeta(p)}  
    \label{eq:1}
\end{equation}
with $\zeta(p)$ deviating from K41 prediction, $p/3$, and being a nonlinear convex function of $p$ with the constraint $\zeta(3)=1$ as imposed by one of the few exact results in HIT, namely the Kolmogorov 4/5 law~\cite{Frisch1995}. The scaling~\eqref{eq:1} is observed within the inertial range, $\eta_K \ll r \ll L_0$, comprised between the Kolmogorov scale, $\eta_K \equiv (\nu^3/\epsilon)^{1/4}$ ($\nu$ being the viscosity and $\epsilon$ the mean rate of energy dissipation) and the large scale $L_0$ of the flow, in HIT coinciding with the forcing scale. The basic idea of the multifractal theory is to assume that, in the inertial range~\cite{Parisi85, Frisch1995},
\begin{equation}
    \label{eq:mf1}
     \delta_r v = U_0 \left(\frac{r}{L_0}\right)^{h} \, ,
\end{equation}
($U_0$  denoting the typical velocity at the scale of energy injection $L_0$) which is consistent with the invariance of the Euler equation under the transformation $r\rightarrow\lambda r$, $v\rightarrow\lambda^h v$ and $t\rightarrow\lambda^{1-h} t$. The K41 theory is a particular instance of \eqref{eq:mf1} with $h=1/3$, consistent with $\zeta(3)=1$.
In the multifractal theory, positive values of $h\le 1$ are allowed with probability 
\begin{equation}
    P_r(h) \sim \left(\frac{r}{L_0}\right)^{3-D(h)} \, .
    \label{eq:mf2}
\end{equation}
Then, using Eqs.~(\ref{eq:mf1})-(\ref{eq:mf2}) and the definition of structure function $S_p(r)=\langle \delta_r v^p\rangle$, the anomalous scaling exponents can be  obtained as the Legendre transform of the function $D(h)$, i.e.,
\begin{equation}
\label{2}
\zeta(p) = \inf_h \{ph +3 -D(h)\}\,.
\end{equation}
One can think of $D(h)$ as the fractal dimension of the set of $h$ or as the large deviation functional describing the statistical properties of $\delta_r v$. The same approach can be followed for the shell model, where the velocity structure functions are defined as $\langle |u_n|^p\rangle \sim k_n^{-\hat{\zeta}_p}$ and display anomalous exponents $\hat{\zeta}_p$ remarkably close to those observed in HIT~\cite{Itamar}. In particular, in this Letter, we assume for $D(h)$ a Log-Poisson functional form~\cite{SL1994}
\begin{equation} 
    D(h) = \frac{3 (h-h_0)}{ \log(\beta)} \left[\log \left( \frac{3(h-h_0)}{d_0 \log(\beta)}\right) -1\right] +3 -d_0 \,.
    \label{eq:SL}
\end{equation}

For the Navier-Stokes equations, a good fit to the anomalous exponents $\zeta(p)$ is given by $h_0=1/9$, $d_0=2$ and $\beta$ fixed by the constraint $\zeta(3)=1$. Notice that the strongest singularity in the flow $h_0$ is characterized by the dimension $D(h_0)=1$, which can be interpreted as the presence of vortex filaments. However, slight variations on these parameters leave practically unaffected the fit, at least for moments that are not too high.
For the Sabra shell model discussed here, a very good fit is obtained with  $h_0=1/9$ and $\beta = 0.6$ (corresponding to $3-d_0=4/3$), as long as we consider values of $\zeta(p)$ for $p$ not too large, see, e.g., Ref.~\cite{de2018time} for a discussion on this point.

\textit{Multifractal model for Lagrangian velocity statistics.}
Similarly to the Eulerian case, we can define the Lagrangian velocity difference over a time interval $\tau$, $\delta_\tau v_L = [v_L(t+\tau) - v_L(t)]$, where $v_L$ is the Lagrangian velocity -- namely, the velocity along a tracer particle trajectory -- and we can define the Lagrangian structure functions $S^L_p(\tau)\equiv \langle |\delta_\tau v_L|^p \rangle$. K41 scaling would suggest that $S^L_p(\tau)\sim \tau^{\xi(p)}$ with $\xi(p)=p/2$ while experiments and simulations~\cite{mordant2001measurement,biferale2008lagrangian} display deviations from the dimensional scaling. Such deviations can be described, similarly to the Eulerian case, with the multifractal model. The first step is to link time and spatial scales, which can be done  via the relation
\begin{equation}
    \tau = \frac{r}{\delta_r v} \sim \frac{L_0}{U_0}\left(\frac{r}{L_0}\right)^{1-h}=T_0\left(\frac{r}{L_0}\right)^{1-h} \, ,
    \label{eq:tau}
\end{equation}
where $T_0$ denotes the large-scale characteristic time. Then,
following the same steps used for Eulerian increments, we can write 
\begin{equation}
    \delta_\tau v_L \sim U_0 \left(\frac{\tau}{T_0}\right)^{\frac{h}{1-h}} \quad \textrm{with} \quad  P_{\tau}(h) \sim \left(\frac{\tau}{T_0}\right)^{\frac{3-D(h)}{1-h}} \, ,
    \label{eq:pditau}
\end{equation}
and compute the anomalous scaling exponents, $\xi(p)$ in the Lagrangian frame as
\begin{eqnarray}
    \label{mfl2}
    \xi(p) = \inf_h \left[ \frac{ph+3-D(h)}{1-h} \right] \, .
\end{eqnarray}
Equations (\ref{2}) and (\ref{mfl2}) show that the same function $D(h)$ can explain the two sets of anomalous scaling exponents $\zeta(p)$ and $\xi(p)$ consistent with the available laboratory and numerical data in HIT~\cite{mordant2001measurement,arneodo2008,benzi2010inertial,benzitoschi}. Similarly, in the shell model the Lagrangian structure functions defined using the velocity in Eq.~\eqref{eq:lagshell} have been shown to display anomalous scaling, $\langle |\delta_\tau v|^p \rangle \sim \tau^{\hat{\xi}(p)}$, with the exponents well captured by the multifractal model~\cite{boffetta2002}. 

In Fig.~\ref{fig:1}(a), we show the kurtosis, $\langle |\delta_\tau v|^4 \rangle/\langle |\delta_\tau v|^2 \rangle^2$, as a function of the time lag $\tau$ for different values of the Reynolds number (\textit{viz.} of the viscosity), demonstrating the strong deviations from Gaussianity due to intermittency and how they increase at decreasing the viscosity. In Fig.~\ref{fig:1}(b), using ideas from extended self-similarity~\cite{benzi1993extended}, we show the local slopes $d\ln(\langle |\delta_\tau v|^4 \rangle)/d\ln(\langle |\delta_\tau v|^2 \rangle)$ vs $\tau$. Similarly to DNS and experimental data~\cite{arneodo2008}, around the Kolmogorov time, we observe a dip, which is usually interpreted as the effect of vortices (see, e.g., \cite{bec2006effects}). 
However, in this case, a well-defined dip is observed even if vortices are absent in the shell model. Moreover, the scaling behavior predicted by the multifractal model (horizontal line) is slowly recovered. The presence of the dip around $\tau_K$ was captured in~\cite{arneodo2008} by using the multifractal model with a supplemental model for the crossover between dissipative and inertial time scale~\cite{chevillard2003}. This crossover is model-dependent (i.e., it depends on whether we are working with the Navier-Stokes equations or the shell model), and we did not attempt to reproduce it for the shell model. However, the above observations are consistent with the fact that the dip originates in the multifractal character of the fluctuations. This observation was indeed first made in~\cite{boffetta2002}, where it was shown that the long crossover disappears when considering a non-intermittent version of the shell model (see Fig.~2 and Fig.~5 and the corresponding discussion in Ref.~\cite{boffetta2002}). It is also worthwhile to remark that machine learning trained on Lagrangian velocity as defined in (\ref{eq:lagshell}) has been successfully applied in predicting and classifying Lagrangian turbulence obtained by DNS data \cite{corbetta2021deep}.

\textit{Multifractal model for the acceleration.}
We now turn to the main focus of this Letter, namely the Lagrangian acceleration $a_L \equiv dv_L/dt$. This quantity displays strong intermittent fluctuations with a long tail in its probability distribution function (PDF) $P(a_L)$. In fact, upon defining $\sigma_a^2 = \langle a_L^2 \rangle$, {\it extreme events} up to several tens of $ \sigma_a$ are usually observed in the time behavior of $a_L$~\cite{la2001fluid,biferale2004}. These extreme events in $a_L$ are typically associated with Lagrangian particles which, along their trajectories, experience the effect of vortex filaments or vortex tubes~\cite{la2001fluid,mordant2004three}. Using the multifractal theory, one can obtain a prediction for the dependence on the Reynolds number $Re$ for both the acceleration moments, $\langle a_L^p\rangle$, and the acceleration PDF, $P(a_L)$. Indeed, we can estimate the Lagrangian acceleration $a_L$ as
\begin{equation} 
    a_L \sim \frac{\delta_{\tau_K} v_L}{\tau_K} \sim \frac{(\delta_{\eta_K} v)^2}{\eta_K}\sim U_0^2 \eta_K^{2h-1} \, ,
    \label{3}
\end{equation}
where the second relation is obtained using \eqref{eq:tau} that links Lagrangian and Eulerian quantities within the multifractal model, and the third using \eqref{eq:mf1}. Then, we observe that, as first suggested by Frisch and Vergassola~\cite{FV1991}, within the multifractal framework, the Kolmogorov scale $\eta_K$ fluctuates as
\begin{equation}  
    \eta_K(h)\sim \left(\frac{\nu L^h}{U_0}\right)^\frac{1}{(1+h)}
    \label{etah}
\end{equation}
as obtained from Eq.~\eqref{eq:mf1} and defining $\eta_K$ with the requirement $(\delta v(\eta_K(h))\eta_K(h))/\nu \sim 1$. Then, using \eqref{eq:mf1} with $r=\eta_K$ in \eqref{3} and \eqref{etah}, recalling that $Re=U_0L_0/\nu$, we end up with
\begin{equation}
    a_L \sim \frac{U^2_0}{L_0} Re^\frac{1-2h}{1+h}
    \label{eq:alRe}
\end{equation}  
with  probability given by \eqref{eq:mf2} for $r=\eta_K(h)$, i.e.,
\begin{equation}
    P(h) \sim \left[\frac{\eta_K(h)}{L} \right]^{3-D(h)} \sim Re^\frac{D(h)-3}{1+h} \, .
    \label{4bis}
\end{equation}
where the last relation is obtained using \eqref{etah}.
Now, from Eqs.~\eqref{eq:alRe} and \eqref{4bis} we get
\begin{equation}
    \langle |a_L|^p\rangle \sim \left(\frac{U_0^2}{L_0}\right)^p Re^{{\alpha}_p} \, , \qquad {\alpha}_p=\sup_{h}\left\{\frac{(1-2h)p-3+D(h)}{1+h} \right\} \, ,
    \label{expacc}
\end{equation}
which, using the log-Poisson model \eqref{eq:SL}, yields a prediction for $\alpha_p$ in excellent agreement with the numerically computed moments within the shell model, as shown in Fig.~\ref{fig:2}.

\begin{figure*}[t!]
\centering
\includegraphics[width=0.5\textwidth]{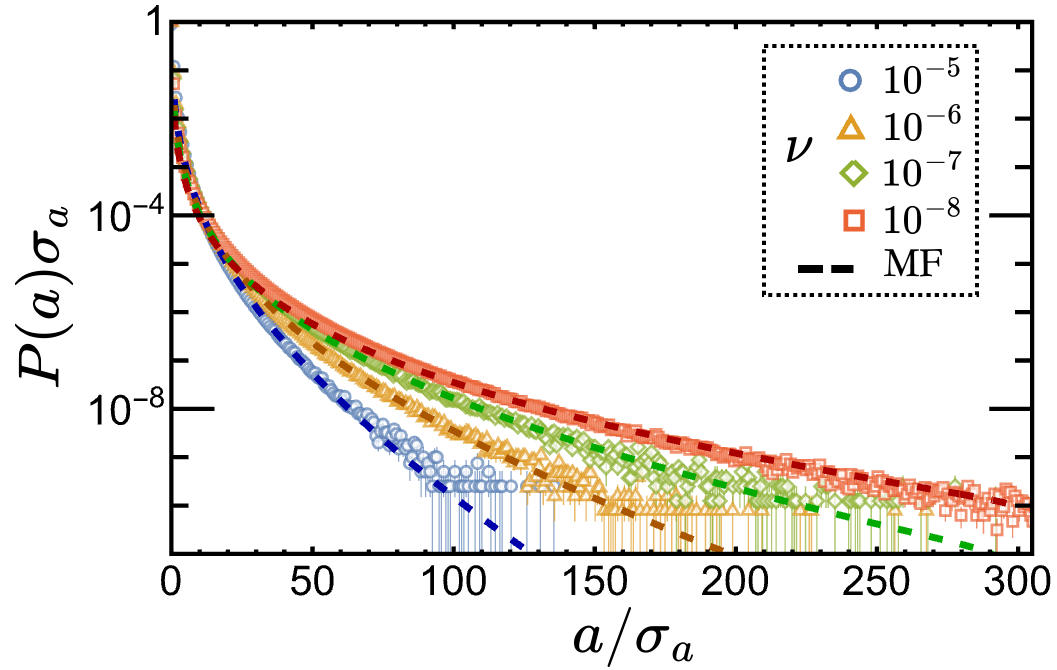}
\caption{\justifying Log-linear plot of the acceleration PDFs normalized with the standard deviation $\sigma_a$ of the acceleration for four different values of the viscosity $\nu=\{10^{-5},10^{-6},10^{-7},10^{-8}\}$, as indicated by the legend. Symbols represent the probability distributions obtained from our numerical simulations, while the corresponding dashed lines of matching color indicate the multifractal prediction.}
\label{fig:3}
\end{figure*}

We now move to the multifractal prediction for the probability density function of the acceleration, $P(a_L)$. We first notice that Eq.~\eqref{3} can be used to express $\eta_K$ in terms of $a_L$. Then combining the expressions (\ref{3}),(\ref{eq:alRe}), and (\ref{4bis}) yields~\cite{biferale2004}
\begin{eqnarray}
    P(a_L) = Z \int {\rm d}h \, a_L^{ \frac{h-5+D(h)}{3} } \nu^ { \frac{ 7-2h-2D(h)}{3}} \exp \left[ -\frac{     a_L^{\frac{2(1+h)}{3}}  \nu^{ \frac{ 2(1-2h)}{3}}  L^{2h}    }{2\sigma^2} \right] \, .
    \label{5}
 \end{eqnarray}
The above equation is derived upon assuming $U_0$ in Eq.~\eqref{eq:mf1} to be a Gaussian variable with zero mean and standard deviation $\sigma$, an extremely good approximation for HIT. While this assumption is not necessary to predict the scaling of the moments, it is essential for modeling the functional form of the PDF.

Following~\cite{biferale2004}, to compare the prediction with numerical data, we can further simplify Eq.~\eqref{5}  by introducing the normalized acceleration $\tilde{a}=a/\sigma_a$ and rewrite it directly for $\tilde{a}$ as
\begin{eqnarray}
  P(\tilde{a}) = Z \int {\rm d}h \, \tilde{a}^{\frac{h-5+D(h)}{3}}
  Re^{\alpha_2 \frac{5-h-D(h)}{6}+\frac{2h+2D(h)-7}{3}}
  \exp \left[ -\frac{     \tilde{a}^{\frac{2(1+h)}{3}}  Re^{\frac{ 2(1-2h)}{3}+\frac{2\alpha_2(1+h)}{3}} }{2} \right] \, ,
    \label{6}
\end{eqnarray}
where $\alpha_2$ is the exponent of the second moment of the acceleration as defined in~\eqref{expacc}.
In the above expression, we used that $\sigma\sim U_0$, i.e., the root mean square velocity is an estimate of the large-scale velocity. The above formula is pleasant as it depends only on one parameter $Re$. Then, using (\ref{eq:SL}) for the $D(h)$, one can numerically compute the above integral and compare the obtained $P(\tilde{a})$ against the numerical findings. It turns out that Eq.~(\ref{6}) provides an excellent qualitative and quantitative prediction of the observed probability distribution of the Lagrangian acceleration, as shown in Fig.~\ref{fig:3}. The curves shown have been obtained by fitting $Re$ for the simulation done with $\nu = 10^{-7}$ and extrapolating the prediction for the others by simply rescaling $Re$ according to the value of $\nu$. Following Ref.~\cite{biferale2004}, the integral in Eq.~\eqref{5} has been computed by setting $h_{min} = 0.16$ to avoid unphysical divergences, and $h_{max} = 0.99$, with the results shown being robust to slight variations of these boundary values. We should note that the normalization of the curves is a free parameter to consider in the fitting procedure, and it may need some fine-tuning. However, this prefactor rescales the curves without changing their shape, and since our focus is on predicting the scaling of the PDF tails, it does not alter our conclusions.
Remarkably, the results shown in Fig.~\ref{fig:3} are qualitatively and quantitatively close to the one observed for the Navier-Stokes equations~\cite{biferale2004}, even if no vortex filaments characterize the intermittent fluctuations of the Sabra model. 

\bigskip

\textit{Conclusions.} 
We have shown that shell models of turbulence, despite lacking coherent vortical structures, reproduce the extreme intermittent fluctuations of Lagrangian acceleration typically observed in real turbulent flows. Using a multifractal framework, we demonstrated that the statistics of acceleration --- including its moments and full probability distribution --- can be accurately predicted across a wide range of Reynolds numbers using only the $D(h)$ extracted from the anomalous scaling exponents $\xi(p)$ without other fitting parameters. 
%These results suggest that the observed intermittency in Lagrangian acceleration stems from inertial-range dynamics rather than specific geometric features such as vortex filaments. 
Building on Refs.~\cite{dombre1998,mailybaev2013}, our findings may be attributed to the existence of statistical coherent structures, manifest in the shell model as a particular phase organization among the velocity modes $u_n$. Crucially, because interactions in shell models are strictly local in scale, any such phase‐organized structures must form through the inertial‐range energy cascade itself, and thus their development is essentially independent of the details of the dissipation mechanism.
Our findings support the universality of multifractal statistics in turbulence and highlight the power of shell models as minimal yet effective tools for capturing key features also of Lagrangian turbulence.

\section*{Acknowledgments}

We thank Luca Biferale for useful discussions. L.P. was supported by the European Research Council (ERC) under the European Union’s Horizon 2020 research and innovation program (Grant Agreement No. 882340).

%\bibliography{biblio}

\begin{thebibliography}{37}%
\makeatletter
\providecommand \@ifxundefined [1]{%
 \@ifx{#1\undefined}
}%
\providecommand \@ifnum [1]{%
 \ifnum #1\expandafter \@firstoftwo
 \else \expandafter \@secondoftwo
 \fi
}%
\providecommand \@ifx [1]{%
 \ifx #1\expandafter \@firstoftwo
 \else \expandafter \@secondoftwo
 \fi
}%
\providecommand \natexlab [1]{#1}%
\providecommand \enquote  [1]{``#1''}%
\providecommand \bibnamefont  [1]{#1}%
\providecommand \bibfnamefont [1]{#1}%
\providecommand \citenamefont [1]{#1}%
\providecommand \href@noop [0]{\@secondoftwo}%
\providecommand \href [0]{\begingroup \@sanitize@url \@href}%
\providecommand \@href[1]{\@@startlink{#1}\@@href}%
\providecommand \@@href[1]{\endgroup#1\@@endlink}%
\providecommand \@sanitize@url [0]{\catcode `\\12\catcode `\$12\catcode
  `\&12\catcode `\#12\catcode `\^12\catcode `\_12\catcode `\%12\relax}%
\providecommand \@@startlink[1]{}%
\providecommand \@@endlink[0]{}%
\providecommand \url  [0]{\begingroup\@sanitize@url \@url }%
\providecommand \@url [1]{\endgroup\@href {#1}{\urlprefix }}%
\providecommand \urlprefix  [0]{URL }%
\providecommand \Eprint [0]{\href }%
\providecommand \doibase [0]{https://doi.org/}%
\providecommand \selectlanguage [0]{\@gobble}%
\providecommand \bibinfo  [0]{\@secondoftwo}%
\providecommand \bibfield  [0]{\@secondoftwo}%
\providecommand \translation [1]{[#1]}%
\providecommand \BibitemOpen [0]{}%
\providecommand \bibitemStop [0]{}%
\providecommand \bibitemNoStop [0]{.\EOS\space}%
\providecommand \EOS [0]{\spacefactor3000\relax}%
\providecommand \BibitemShut  [1]{\csname bibitem#1\endcsname}%
\let\auto@bib@innerbib\@empty
%</preamble>
\bibitem [{\citenamefont {Kolmogorov}(1941)}]{Kolmogorov41}%
  \BibitemOpen
  \bibfield  {author} {\bibinfo {author} {\bibfnamefont {A.~N.}\ \bibnamefont
  {Kolmogorov}},\ }\bibfield  {title} {\bibinfo {title} {{The local structure
  of turbulence in incompressible viscous fluid for very large {R}eynold
  number}},\ }\href@noop {} {\bibfield  {journal} {\bibinfo  {journal} {Dokl.
  Akad. Nauk. SSSR}\ }\textbf {\bibinfo {volume} {30}},\ \bibinfo {pages} {299}
  (\bibinfo {year} {1941})}\BibitemShut {NoStop}%
\bibitem [{\citenamefont {Frisch}(1995)}]{Frisch1995}%
  \BibitemOpen
  \bibfield  {author} {\bibinfo {author} {\bibfnamefont {U.}~\bibnamefont
  {Frisch}},\ }\href@noop {} {\emph {\bibinfo {title} {Turbulence: the legacy
  of AN Kolmogorov}}}\ (\bibinfo  {publisher} {Cambridge University Press},\
  \bibinfo {year} {1995})\BibitemShut {NoStop}%
\bibitem [{\citenamefont {Arneodo}\ \emph {et~al.}(1996)\citenamefont
  {Arneodo}, \citenamefont {Baudet}, \citenamefont {Belin}, \citenamefont
  {Benzi}, \citenamefont {Castaing}, \citenamefont {Chabaud}, \citenamefont
  {Chavarria}, \citenamefont {Ciliberto}, \citenamefont {Camussi},
  \citenamefont {Chillà}, \citenamefont {Dubrulle}, \citenamefont {Gagne},
  \citenamefont {Hebral}, \citenamefont {Herweijer}, \citenamefont {Marchand},
  \citenamefont {Maurer}, \citenamefont {Muzy}, \citenamefont {Naert},
  \citenamefont {Noullez}, \citenamefont {Peinke}, \citenamefont {Roux},
  \citenamefont {Tabeling}, \citenamefont {van~de Water},\ and\ \citenamefont
  {Willaime}}]{arneodo1996}%
  \BibitemOpen
  \bibfield  {author} {\bibinfo {author} {\bibfnamefont {A.}~\bibnamefont
  {Arneodo}}, \bibinfo {author} {\bibfnamefont {C.}~\bibnamefont {Baudet}},
  \bibinfo {author} {\bibfnamefont {F.}~\bibnamefont {Belin}}, \bibinfo
  {author} {\bibfnamefont {R.}~\bibnamefont {Benzi}}, \bibinfo {author}
  {\bibfnamefont {B.}~\bibnamefont {Castaing}}, \bibinfo {author}
  {\bibfnamefont {B.}~\bibnamefont {Chabaud}}, \bibinfo {author} {\bibfnamefont
  {R.}~\bibnamefont {Chavarria}}, \bibinfo {author} {\bibfnamefont
  {S.}~\bibnamefont {Ciliberto}}, \bibinfo {author} {\bibfnamefont
  {R.}~\bibnamefont {Camussi}}, \bibinfo {author} {\bibfnamefont
  {F.}~\bibnamefont {Chillà}}, \bibinfo {author} {\bibfnamefont
  {B.}~\bibnamefont {Dubrulle}}, \bibinfo {author} {\bibfnamefont
  {Y.}~\bibnamefont {Gagne}}, \bibinfo {author} {\bibfnamefont
  {B.}~\bibnamefont {Hebral}}, \bibinfo {author} {\bibfnamefont
  {J.}~\bibnamefont {Herweijer}}, \bibinfo {author} {\bibfnamefont
  {M.}~\bibnamefont {Marchand}}, \bibinfo {author} {\bibfnamefont
  {J.}~\bibnamefont {Maurer}}, \bibinfo {author} {\bibfnamefont {J.~F.}\
  \bibnamefont {Muzy}}, \bibinfo {author} {\bibfnamefont {A.}~\bibnamefont
  {Naert}}, \bibinfo {author} {\bibfnamefont {A.}~\bibnamefont {Noullez}},
  \bibinfo {author} {\bibfnamefont {J.}~\bibnamefont {Peinke}}, \bibinfo
  {author} {\bibfnamefont {F.}~\bibnamefont {Roux}}, \bibinfo {author}
  {\bibfnamefont {P.}~\bibnamefont {Tabeling}}, \bibinfo {author}
  {\bibfnamefont {W.}~\bibnamefont {van~de Water}},\ and\ \bibinfo {author}
  {\bibfnamefont {H.}~\bibnamefont {Willaime}},\ }\bibfield  {title} {\bibinfo
  {title} {Structure functions in turbulence, in various flow configurations,
  at reynolds number between 30 and 5000, using extended self-similarity},\
  }\href@noop {} {\bibfield  {journal} {\bibinfo  {journal} {EPL}\ }\textbf
  {\bibinfo {volume} {34}},\ \bibinfo {pages} {411} (\bibinfo {year}
  {1996})}\BibitemShut {NoStop}%
\bibitem [{\citenamefont {Anselmet}\ \emph {et~al.}(2001)\citenamefont
  {Anselmet}, \citenamefont {Antonia},\ and\ \citenamefont
  {Danaila}}]{anselmet2001turbulent}%
  \BibitemOpen
  \bibfield  {author} {\bibinfo {author} {\bibfnamefont {F.}~\bibnamefont
  {Anselmet}}, \bibinfo {author} {\bibfnamefont {R.}~\bibnamefont {Antonia}},\
  and\ \bibinfo {author} {\bibfnamefont {L.}~\bibnamefont {Danaila}},\
  }\bibfield  {title} {\bibinfo {title} {Turbulent flows and intermittency in
  laboratory experiments},\ }\href@noop {} {\bibfield  {journal} {\bibinfo
  {journal} {Planet. Space Sci.}\ }\textbf {\bibinfo {volume} {49}},\ \bibinfo
  {pages} {1177} (\bibinfo {year} {2001})}\BibitemShut {NoStop}%
\bibitem [{\citenamefont {Kraichnan}(1968)}]{kraichnan1968small}%
  \BibitemOpen
  \bibfield  {author} {\bibinfo {author} {\bibfnamefont {R.~H.}\ \bibnamefont
  {Kraichnan}},\ }\bibfield  {title} {\bibinfo {title} {Small-scale structure
  of a scalar field convected by turbulence},\ }\href@noop {} {\bibfield
  {journal} {\bibinfo  {journal} {Phys. Fluids}\ }\textbf {\bibinfo {volume}
  {11}},\ \bibinfo {pages} {945} (\bibinfo {year} {1968})}\BibitemShut
  {NoStop}%
\bibitem [{\citenamefont {Gawedzki}\ and\ \citenamefont
  {Kupiainen}(1995)}]{gawedzki1995anomalous}%
  \BibitemOpen
  \bibfield  {author} {\bibinfo {author} {\bibfnamefont {K.}~\bibnamefont
  {Gawedzki}}\ and\ \bibinfo {author} {\bibfnamefont {A.}~\bibnamefont
  {Kupiainen}},\ }\bibfield  {title} {\bibinfo {title} {Anomalous scaling of
  the passive scalar},\ }\href@noop {} {\bibfield  {journal} {\bibinfo
  {journal} {Phys. Rev. Lett.}\ }\textbf {\bibinfo {volume} {75}},\ \bibinfo
  {pages} {3834} (\bibinfo {year} {1995})}\BibitemShut {NoStop}%
\bibitem [{\citenamefont {Chertkov}\ and\ \citenamefont
  {Falkovich}(1996)}]{chertkov1996anomalous}%
  \BibitemOpen
  \bibfield  {author} {\bibinfo {author} {\bibfnamefont {M.}~\bibnamefont
  {Chertkov}}\ and\ \bibinfo {author} {\bibfnamefont {G.}~\bibnamefont
  {Falkovich}},\ }\bibfield  {title} {\bibinfo {title} {Anomalous scaling
  exponents of a white-advected passive scalar},\ }\href@noop {} {\bibfield
  {journal} {\bibinfo  {journal} {Phys. Rev. Lett.}\ }\textbf {\bibinfo
  {volume} {76}},\ \bibinfo {pages} {2706} (\bibinfo {year}
  {1996})}\BibitemShut {NoStop}%
\bibitem [{\citenamefont {Shraiman}\ and\ \citenamefont
  {Siggia}(1996)}]{shraiman1996symmetry}%
  \BibitemOpen
  \bibfield  {author} {\bibinfo {author} {\bibfnamefont {B.~I.}\ \bibnamefont
  {Shraiman}}\ and\ \bibinfo {author} {\bibfnamefont {E.~D.}\ \bibnamefont
  {Siggia}},\ }\bibfield  {title} {\bibinfo {title} {Symmetry and scaling of
  turbulent mixing},\ }\href@noop {} {\bibfield  {journal} {\bibinfo  {journal}
  {Phys. Rev. Lett.}\ }\textbf {\bibinfo {volume} {77}},\ \bibinfo {pages}
  {2463} (\bibinfo {year} {1996})}\BibitemShut {NoStop}%
\bibitem [{\citenamefont {Frisch}\ \emph {et~al.}(1999)\citenamefont {Frisch},
  \citenamefont {Mazzino}, \citenamefont {Noullez},\ and\ \citenamefont
  {Vergassola}}]{frisch1999lagrangian}%
  \BibitemOpen
  \bibfield  {author} {\bibinfo {author} {\bibfnamefont {U.}~\bibnamefont
  {Frisch}}, \bibinfo {author} {\bibfnamefont {A.}~\bibnamefont {Mazzino}},
  \bibinfo {author} {\bibfnamefont {A.}~\bibnamefont {Noullez}},\ and\ \bibinfo
  {author} {\bibfnamefont {M.}~\bibnamefont {Vergassola}},\ }\bibfield  {title}
  {\bibinfo {title} {{L}agrangian method for multiple correlations in passive
  scalar advection},\ }\href@noop {} {\bibfield  {journal} {\bibinfo  {journal}
  {Phys. Fluids}\ }\textbf {\bibinfo {volume} {11}},\ \bibinfo {pages} {2178}
  (\bibinfo {year} {1999})}\BibitemShut {NoStop}%
\bibitem [{\citenamefont {Shraiman}\ and\ \citenamefont
  {Siggia}(2000)}]{shraiman2000scalar}%
  \BibitemOpen
  \bibfield  {author} {\bibinfo {author} {\bibfnamefont {B.~I.}\ \bibnamefont
  {Shraiman}}\ and\ \bibinfo {author} {\bibfnamefont {E.~D.}\ \bibnamefont
  {Siggia}},\ }\bibfield  {title} {\bibinfo {title} {Scalar turbulence},\
  }\href@noop {} {\bibfield  {journal} {\bibinfo  {journal} {Nature}\ }\textbf
  {\bibinfo {volume} {405}},\ \bibinfo {pages} {639} (\bibinfo {year}
  {2000})}\BibitemShut {NoStop}%
\bibitem [{\citenamefont {Falkovich}\ \emph {et~al.}(2001)\citenamefont
  {Falkovich}, \citenamefont {Gawedzki},\ and\ \citenamefont
  {Vergassola}}]{falkovich2001particles}%
  \BibitemOpen
  \bibfield  {author} {\bibinfo {author} {\bibfnamefont {G.}~\bibnamefont
  {Falkovich}}, \bibinfo {author} {\bibfnamefont {K.}~\bibnamefont
  {Gawedzki}},\ and\ \bibinfo {author} {\bibfnamefont {M.}~\bibnamefont
  {Vergassola}},\ }\bibfield  {title} {\bibinfo {title} {Particles and fields
  in fluid turbulence},\ }\href@noop {} {\bibfield  {journal} {\bibinfo
  {journal} {Rev. Mod. Phys.}\ }\textbf {\bibinfo {volume} {73}},\ \bibinfo
  {pages} {913} (\bibinfo {year} {2001})}\BibitemShut {NoStop}%
\bibitem [{\citenamefont {Parisi}\ and\ \citenamefont
  {Frisch}(1985)}]{Parisi85}%
  \BibitemOpen
  \bibfield  {author} {\bibinfo {author} {\bibfnamefont {G.}~\bibnamefont
  {Parisi}}\ and\ \bibinfo {author} {\bibfnamefont {U.}~\bibnamefont
  {Frisch}},\ }\bibfield  {title} {\bibinfo {title} {{On the singularity
  structure of fully developed turbulence}},\ }in\ \href@noop {} {\emph
  {\bibinfo {booktitle} {{Turbulence and predictability of geophysical fluid
  dynamics}}}},\ \bibinfo {editor} {edited by\ \bibinfo {editor} {\bibfnamefont
  {M.}~\bibnamefont {Ghil}}, \bibinfo {editor} {\bibfnamefont {G.}~\bibnamefont
  {Parisi}},\ and\ \bibinfo {editor} {\bibfnamefont {R.}~\bibnamefont
  {Benzi}}}\ (\bibinfo  {publisher} {North-Holland},\ \bibinfo {address}
  {Amsterdam},\ \bibinfo {year} {1985})\ p.~\bibinfo {pages} {84}\BibitemShut
  {NoStop}%
\bibitem [{\citenamefont {Mordant}\ \emph {et~al.}(2001)\citenamefont
  {Mordant}, \citenamefont {Metz}, \citenamefont {Michel},\ and\ \citenamefont
  {Pinton}}]{mordant2001measurement}%
  \BibitemOpen
  \bibfield  {author} {\bibinfo {author} {\bibfnamefont {N.}~\bibnamefont
  {Mordant}}, \bibinfo {author} {\bibfnamefont {P.}~\bibnamefont {Metz}},
  \bibinfo {author} {\bibfnamefont {O.}~\bibnamefont {Michel}},\ and\ \bibinfo
  {author} {\bibfnamefont {J.-F.}\ \bibnamefont {Pinton}},\ }\bibfield  {title}
  {\bibinfo {title} {Measurement of {L}agrangian velocity in fully developed
  turbulence},\ }\href@noop {} {\bibfield  {journal} {\bibinfo  {journal}
  {Phys. Rev. Lett.}\ }\textbf {\bibinfo {volume} {87}},\ \bibinfo {pages}
  {214501} (\bibinfo {year} {2001})}\BibitemShut {NoStop}%
\bibitem [{\citenamefont {Boffetta}\ \emph {et~al.}(2002)\citenamefont
  {Boffetta}, \citenamefont {De~Lillo},\ and\ \citenamefont
  {Musacchio}}]{boffetta2002}%
  \BibitemOpen
  \bibfield  {author} {\bibinfo {author} {\bibfnamefont {G.}~\bibnamefont
  {Boffetta}}, \bibinfo {author} {\bibfnamefont {F.}~\bibnamefont {De~Lillo}},\
  and\ \bibinfo {author} {\bibfnamefont {S.}~\bibnamefont {Musacchio}},\
  }\bibfield  {title} {\bibinfo {title} {{L}agrangian statistics and temporal
  intermittency in a shell model of turbulence},\ }\href@noop {} {\bibfield
  {journal} {\bibinfo  {journal} {Phys. Rev. E}\ }\textbf {\bibinfo {volume}
  {66}},\ \bibinfo {pages} {066307} (\bibinfo {year} {2002})}\BibitemShut
  {NoStop}%
\bibitem [{\citenamefont {Arn{\'e}odo}\ \emph {et~al.}(2008)\citenamefont
  {Arn{\'e}odo} \emph {et~al.}}]{arneodo2008}%
  \BibitemOpen
  \bibfield  {author} {\bibinfo {author} {\bibfnamefont {A.}~\bibnamefont
  {Arn{\'e}odo}} \emph {et~al.},\ }\bibfield  {title} {\bibinfo {title}
  {Universal intermittent properties of particle trajectories in highly
  turbulent flows},\ }\href@noop {} {\bibfield  {journal} {\bibinfo  {journal}
  {Phys. Rev. Lett.}\ }\textbf {\bibinfo {volume} {100}},\ \bibinfo {pages}
  {254504} (\bibinfo {year} {2008})}\BibitemShut {NoStop}%
\bibitem [{\citenamefont {La~Porta}\ \emph {et~al.}(2001)\citenamefont
  {La~Porta}, \citenamefont {Voth}, \citenamefont {Crawford}, \citenamefont
  {Alexander},\ and\ \citenamefont {Bodenschatz}}]{la2001fluid}%
  \BibitemOpen
  \bibfield  {author} {\bibinfo {author} {\bibfnamefont {A.}~\bibnamefont
  {La~Porta}}, \bibinfo {author} {\bibfnamefont {G.~A.}\ \bibnamefont {Voth}},
  \bibinfo {author} {\bibfnamefont {A.~M.}\ \bibnamefont {Crawford}}, \bibinfo
  {author} {\bibfnamefont {J.}~\bibnamefont {Alexander}},\ and\ \bibinfo
  {author} {\bibfnamefont {E.}~\bibnamefont {Bodenschatz}},\ }\bibfield
  {title} {\bibinfo {title} {Fluid particle accelerations in fully developed
  turbulence},\ }\href@noop {} {\bibfield  {journal} {\bibinfo  {journal}
  {Nature}\ }\textbf {\bibinfo {volume} {409}},\ \bibinfo {pages} {1017}
  (\bibinfo {year} {2001})}\BibitemShut {NoStop}%
\bibitem [{\citenamefont {Mordant}\ \emph {et~al.}(2004)\citenamefont
  {Mordant}, \citenamefont {Crawford},\ and\ \citenamefont
  {Bodenschatz}}]{mordant2004three}%
  \BibitemOpen
  \bibfield  {author} {\bibinfo {author} {\bibfnamefont {N.}~\bibnamefont
  {Mordant}}, \bibinfo {author} {\bibfnamefont {A.~M.}\ \bibnamefont
  {Crawford}},\ and\ \bibinfo {author} {\bibfnamefont {E.}~\bibnamefont
  {Bodenschatz}},\ }\bibfield  {title} {\bibinfo {title} {Three-dimensional
  structure of the {L}agrangian acceleration in turbulent flows},\ }\href@noop
  {} {\bibfield  {journal} {\bibinfo  {journal} {Phys. Rev. Lett.}\ }\textbf
  {\bibinfo {volume} {93}},\ \bibinfo {pages} {214501} (\bibinfo {year}
  {2004})}\BibitemShut {NoStop}%
\bibitem [{\citenamefont {Biferale}\ \emph {et~al.}(2004)\citenamefont
  {Biferale}, \citenamefont {Boffetta}, \citenamefont {Celani}, \citenamefont
  {Devenish}, \citenamefont {Lanotte},\ and\ \citenamefont
  {Toschi}}]{biferale2004}%
  \BibitemOpen
  \bibfield  {author} {\bibinfo {author} {\bibfnamefont {L.}~\bibnamefont
  {Biferale}}, \bibinfo {author} {\bibfnamefont {G.}~\bibnamefont {Boffetta}},
  \bibinfo {author} {\bibfnamefont {A.}~\bibnamefont {Celani}}, \bibinfo
  {author} {\bibfnamefont {B.~J.}\ \bibnamefont {Devenish}}, \bibinfo {author}
  {\bibfnamefont {A.~S.}\ \bibnamefont {Lanotte}},\ and\ \bibinfo {author}
  {\bibfnamefont {F.}~\bibnamefont {Toschi}},\ }\bibfield  {title} {\bibinfo
  {title} {Multifractal statistics of {L}agrangian velocity and acceleration in
  turbulence},\ }\href@noop {} {\bibfield  {journal} {\bibinfo  {journal}
  {Phys. Rev. Lett.}\ }\textbf {\bibinfo {volume} {93}},\ \bibinfo {pages}
  {064502} (\bibinfo {year} {2004})}\BibitemShut {NoStop}%
\bibitem [{\citenamefont {Buaria}\ and\ \citenamefont
  {Sreenivasan}(2022)}]{buaria2022}%
  \BibitemOpen
  \bibfield  {author} {\bibinfo {author} {\bibfnamefont {D.}~\bibnamefont
  {Buaria}}\ and\ \bibinfo {author} {\bibfnamefont {K.~R.}\ \bibnamefont
  {Sreenivasan}},\ }\bibfield  {title} {\bibinfo {title} {Scaling of
  acceleration statistics in high {R}eynolds number turbulence},\ }\href
  {https://doi.org/10.1103/PhysRevLett.128.234502} {\bibfield  {journal}
  {\bibinfo  {journal} {Phys. Rev. Lett.}\ }\textbf {\bibinfo {volume} {128}},\
  \bibinfo {pages} {234502} (\bibinfo {year} {2022})}\BibitemShut {NoStop}%
\bibitem [{\citenamefont {Biferale}\ \emph {et~al.}(2005)\citenamefont
  {Biferale}, \citenamefont {Boffetta}, \citenamefont {Celani}, \citenamefont
  {Lanotte},\ and\ \citenamefont {Toschi}}]{biferale2005particle}%
  \BibitemOpen
  \bibfield  {author} {\bibinfo {author} {\bibfnamefont {L.}~\bibnamefont
  {Biferale}}, \bibinfo {author} {\bibfnamefont {G.}~\bibnamefont {Boffetta}},
  \bibinfo {author} {\bibfnamefont {A.}~\bibnamefont {Celani}}, \bibinfo
  {author} {\bibfnamefont {A.}~\bibnamefont {Lanotte}},\ and\ \bibinfo {author}
  {\bibfnamefont {F.}~\bibnamefont {Toschi}},\ }\bibfield  {title} {\bibinfo
  {title} {Particle trapping in three-dimensional fully developed turbulence},\
  }\href@noop {} {\bibfield  {journal} {\bibinfo  {journal} {Phys. Fluids}\
  }\textbf {\bibinfo {volume} {17}} (\bibinfo {year} {2005})}\BibitemShut
  {NoStop}%
\bibitem [{\citenamefont {Frisch}\ and\ \citenamefont
  {Vergassola}(1991)}]{FV1991}%
  \BibitemOpen
  \bibfield  {author} {\bibinfo {author} {\bibfnamefont {U.}~\bibnamefont
  {Frisch}}\ and\ \bibinfo {author} {\bibfnamefont {M.}~\bibnamefont
  {Vergassola}},\ }\bibfield  {title} {\bibinfo {title} {A prediction of the
  multifractal model: the intermediate dissipation range},\ }\href@noop {}
  {\bibfield  {journal} {\bibinfo  {journal} {EPL}\ }\textbf {\bibinfo {volume}
  {14}},\ \bibinfo {pages} {439} (\bibinfo {year} {1991})}\BibitemShut
  {NoStop}%
\bibitem [{\citenamefont {Biferale}\ \emph {et~al.}(1999)\citenamefont
  {Biferale}, \citenamefont {Cencini}, \citenamefont {Vergni},\ and\
  \citenamefont {Vulpiani}}]{biferale1999exit}%
  \BibitemOpen
  \bibfield  {author} {\bibinfo {author} {\bibfnamefont {L.}~\bibnamefont
  {Biferale}}, \bibinfo {author} {\bibfnamefont {M.}~\bibnamefont {Cencini}},
  \bibinfo {author} {\bibfnamefont {D.}~\bibnamefont {Vergni}},\ and\ \bibinfo
  {author} {\bibfnamefont {A.}~\bibnamefont {Vulpiani}},\ }\bibfield  {title}
  {\bibinfo {title} {Exit time of turbulent signals: A way to detect the
  intermediate dissipative range},\ }\href@noop {} {\bibfield  {journal}
  {\bibinfo  {journal} {Phys. Rev. E}\ }\textbf {\bibinfo {volume} {60}},\
  \bibinfo {pages} {R6295} (\bibinfo {year} {1999})}\BibitemShut {NoStop}%
\bibitem [{\citenamefont {Boffetta}\ \emph {et~al.}(2008)\citenamefont
  {Boffetta}, \citenamefont {Mazzino},\ and\ \citenamefont
  {Vulpiani}}]{boffetta2008twenty}%
  \BibitemOpen
  \bibfield  {author} {\bibinfo {author} {\bibfnamefont {G.}~\bibnamefont
  {Boffetta}}, \bibinfo {author} {\bibfnamefont {A.}~\bibnamefont {Mazzino}},\
  and\ \bibinfo {author} {\bibfnamefont {A.}~\bibnamefont {Vulpiani}},\
  }\bibfield  {title} {\bibinfo {title} {Twenty-five years of multifractals in
  fully developed turbulence: a tribute to {G}iovanni {P}aladin},\ }\href@noop
  {} {\bibfield  {journal} {\bibinfo  {journal} {J. Phys. A}\ }\textbf
  {\bibinfo {volume} {41}},\ \bibinfo {pages} {363001} (\bibinfo {year}
  {2008})}\BibitemShut {NoStop}%
\bibitem [{\citenamefont {Biferale}(2003)}]{biferale2003}%
  \BibitemOpen
  \bibfield  {author} {\bibinfo {author} {\bibfnamefont {L.}~\bibnamefont
  {Biferale}},\ }\bibfield  {title} {\bibinfo {title} {Shell models of energy
  cascade in turbulence},\ }\href@noop {} {\bibfield  {journal} {\bibinfo
  {journal} {Annu. Rev. Fluid Mech.}\ }\textbf {\bibinfo {volume} {35}},\
  \bibinfo {pages} {441} (\bibinfo {year} {2003})}\BibitemShut {NoStop}%
\bibitem [{\citenamefont {Bohr}\ \emph {et~al.}(2005)\citenamefont {Bohr},
  \citenamefont {Jensen}, \citenamefont {Paladin},\ and\ \citenamefont
  {Vulpiani}}]{bohr2005}%
  \BibitemOpen
  \bibfield  {author} {\bibinfo {author} {\bibfnamefont {T.}~\bibnamefont
  {Bohr}}, \bibinfo {author} {\bibfnamefont {M.~H.}\ \bibnamefont {Jensen}},
  \bibinfo {author} {\bibfnamefont {G.}~\bibnamefont {Paladin}},\ and\ \bibinfo
  {author} {\bibfnamefont {A.}~\bibnamefont {Vulpiani}},\ }\href@noop {} {\emph
  {\bibinfo {title} {Dynamical systems approach to turbulence}}}\ (\bibinfo
  {publisher} {Cambridge University Press},\ \bibinfo {year}
  {2005})\BibitemShut {NoStop}%
\bibitem [{\citenamefont {L’vov}\ \emph {et~al.}(1998)\citenamefont
  {L’vov}, \citenamefont {Podivilov}, \citenamefont {Pomyalov}, \citenamefont
  {Procaccia},\ and\ \citenamefont {Vandembroucq}}]{Itamar}%
  \BibitemOpen
  \bibfield  {author} {\bibinfo {author} {\bibfnamefont {V.~S.}\ \bibnamefont
  {L’vov}}, \bibinfo {author} {\bibfnamefont {E.}~\bibnamefont {Podivilov}},
  \bibinfo {author} {\bibfnamefont {A.}~\bibnamefont {Pomyalov}}, \bibinfo
  {author} {\bibfnamefont {I.}~\bibnamefont {Procaccia}},\ and\ \bibinfo
  {author} {\bibfnamefont {D.}~\bibnamefont {Vandembroucq}},\ }\bibfield
  {title} {\bibinfo {title} {Improved shell model of turbulence},\ }\href@noop
  {} {\bibfield  {journal} {\bibinfo  {journal} {Phys. Rev. E}\ }\textbf
  {\bibinfo {volume} {58}},\ \bibinfo {pages} {1811} (\bibinfo {year}
  {1998})}\BibitemShut {NoStop}%
\bibitem [{\citenamefont {Benzi}\ and\ \citenamefont
  {Toschi}(2023)}]{benzitoschi}%
  \BibitemOpen
  \bibfield  {author} {\bibinfo {author} {\bibfnamefont {R.}~\bibnamefont
  {Benzi}}\ and\ \bibinfo {author} {\bibfnamefont {F.}~\bibnamefont {Toschi}},\
  }\bibfield  {title} {\bibinfo {title} {Lectures on turbulence},\ }\href@noop
  {} {\bibfield  {journal} {\bibinfo  {journal} {Phys. Rep.}\ }\textbf
  {\bibinfo {volume} {1021}},\ \bibinfo {pages} {1} (\bibinfo {year}
  {2023})}\BibitemShut {NoStop}%
\bibitem [{\citenamefont {She}\ and\ \citenamefont {Leveque}(1994)}]{SL1994}%
  \BibitemOpen
  \bibfield  {author} {\bibinfo {author} {\bibfnamefont {Z.-S.}\ \bibnamefont
  {She}}\ and\ \bibinfo {author} {\bibfnamefont {E.}~\bibnamefont {Leveque}},\
  }\bibfield  {title} {\bibinfo {title} {Universal scaling laws in fully
  developed turbulence},\ }\href@noop {} {\bibfield  {journal} {\bibinfo
  {journal} {Phys. Rev. Lett.}\ }\textbf {\bibinfo {volume} {72}},\ \bibinfo
  {pages} {336} (\bibinfo {year} {1994})}\BibitemShut {NoStop}%
\bibitem [{\citenamefont {De~Pietro}\ \emph {et~al.}(2018)\citenamefont
  {De~Pietro}, \citenamefont {Biferale}, \citenamefont {Boffetta},\ and\
  \citenamefont {Cencini}}]{de2018time}%
  \BibitemOpen
  \bibfield  {author} {\bibinfo {author} {\bibfnamefont {M.}~\bibnamefont
  {De~Pietro}}, \bibinfo {author} {\bibfnamefont {L.}~\bibnamefont {Biferale}},
  \bibinfo {author} {\bibfnamefont {G.}~\bibnamefont {Boffetta}},\ and\
  \bibinfo {author} {\bibfnamefont {M.}~\bibnamefont {Cencini}},\ }\bibfield
  {title} {\bibinfo {title} {Time irreversibility in reversible shell models of
  turbulence},\ }\href@noop {} {\bibfield  {journal} {\bibinfo  {journal} {Eur.
  Phys. J. E}\ }\textbf {\bibinfo {volume} {41}},\ \bibinfo {pages} {1}
  (\bibinfo {year} {2018})}\BibitemShut {NoStop}%
\bibitem [{\citenamefont {Biferale}\ \emph {et~al.}(2008)\citenamefont
  {Biferale}, \citenamefont {Bodenschatz}, \citenamefont {Cencini},
  \citenamefont {Lanotte}, \citenamefont {Ouellette}, \citenamefont {Toschi},\
  and\ \citenamefont {Xu}}]{biferale2008lagrangian}%
  \BibitemOpen
  \bibfield  {author} {\bibinfo {author} {\bibfnamefont {L.}~\bibnamefont
  {Biferale}}, \bibinfo {author} {\bibfnamefont {E.}~\bibnamefont
  {Bodenschatz}}, \bibinfo {author} {\bibfnamefont {M.}~\bibnamefont
  {Cencini}}, \bibinfo {author} {\bibfnamefont {A.~S.}\ \bibnamefont
  {Lanotte}}, \bibinfo {author} {\bibfnamefont {N.~T.}\ \bibnamefont
  {Ouellette}}, \bibinfo {author} {\bibfnamefont {F.}~\bibnamefont {Toschi}},\
  and\ \bibinfo {author} {\bibfnamefont {H.}~\bibnamefont {Xu}},\ }\bibfield
  {title} {\bibinfo {title} {{L}agrangian structure functions in turbulence: A
  quantitative comparison between experiment and direct numerical simulation},\
  }\href@noop {} {\bibfield  {journal} {\bibinfo  {journal} {Phys. Fluids}\
  }\textbf {\bibinfo {volume} {20}} (\bibinfo {year} {2008})}\BibitemShut
  {NoStop}%
\bibitem [{\citenamefont {Benzi}\ \emph {et~al.}(2010)\citenamefont {Benzi},
  \citenamefont {Biferale}, \citenamefont {Fisher}, \citenamefont {Lamb},\ and\
  \citenamefont {Toschi}}]{benzi2010inertial}%
  \BibitemOpen
  \bibfield  {author} {\bibinfo {author} {\bibfnamefont {R.}~\bibnamefont
  {Benzi}}, \bibinfo {author} {\bibfnamefont {L.}~\bibnamefont {Biferale}},
  \bibinfo {author} {\bibfnamefont {R.}~\bibnamefont {Fisher}}, \bibinfo
  {author} {\bibfnamefont {D.}~\bibnamefont {Lamb}},\ and\ \bibinfo {author}
  {\bibfnamefont {F.}~\bibnamefont {Toschi}},\ }\bibfield  {title} {\bibinfo
  {title} {Inertial range {E}ulerian and {L}agrangian statistics from numerical
  simulations of isotropic turbulence},\ }\href@noop {} {\bibfield  {journal}
  {\bibinfo  {journal} {J. Fluid Mech.}\ }\textbf {\bibinfo {volume} {653}},\
  \bibinfo {pages} {221} (\bibinfo {year} {2010})}\BibitemShut {NoStop}%
\bibitem [{\citenamefont {Benzi}\ \emph {et~al.}(1993)\citenamefont {Benzi},
  \citenamefont {Ciliberto}, \citenamefont {Tripiccione}, \citenamefont
  {Baudet}, \citenamefont {Massaioli},\ and\ \citenamefont
  {Succi}}]{benzi1993extended}%
  \BibitemOpen
  \bibfield  {author} {\bibinfo {author} {\bibfnamefont {R.}~\bibnamefont
  {Benzi}}, \bibinfo {author} {\bibfnamefont {S.}~\bibnamefont {Ciliberto}},
  \bibinfo {author} {\bibfnamefont {R.}~\bibnamefont {Tripiccione}}, \bibinfo
  {author} {\bibfnamefont {C.}~\bibnamefont {Baudet}}, \bibinfo {author}
  {\bibfnamefont {F.}~\bibnamefont {Massaioli}},\ and\ \bibinfo {author}
  {\bibfnamefont {S.}~\bibnamefont {Succi}},\ }\bibfield  {title} {\bibinfo
  {title} {Extended self-similarity in turbulent flows},\ }\href@noop {}
  {\bibfield  {journal} {\bibinfo  {journal} {Phys. Rev. E}\ }\textbf {\bibinfo
  {volume} {48}},\ \bibinfo {pages} {R29} (\bibinfo {year} {1993})}\BibitemShut
  {NoStop}%
\bibitem [{\citenamefont {Bec}\ \emph {et~al.}(2006)\citenamefont {Bec},
  \citenamefont {Biferale}, \citenamefont {Cencini}, \citenamefont {Lanotte},\
  and\ \citenamefont {Toschi}}]{bec2006effects}%
  \BibitemOpen
  \bibfield  {author} {\bibinfo {author} {\bibfnamefont {J.}~\bibnamefont
  {Bec}}, \bibinfo {author} {\bibfnamefont {L.}~\bibnamefont {Biferale}},
  \bibinfo {author} {\bibfnamefont {M.}~\bibnamefont {Cencini}}, \bibinfo
  {author} {\bibfnamefont {A.~S.}\ \bibnamefont {Lanotte}},\ and\ \bibinfo
  {author} {\bibfnamefont {F.}~\bibnamefont {Toschi}},\ }\bibfield  {title}
  {\bibinfo {title} {Effects of vortex filaments on the velocity of tracers and
  heavy particles in turbulence},\ }\href@noop {} {\bibfield  {journal}
  {\bibinfo  {journal} {Phys. Fluids}\ }\textbf {\bibinfo {volume} {18}}
  (\bibinfo {year} {2006})}\BibitemShut {NoStop}%
\bibitem [{\citenamefont {Chevillard}\ \emph {et~al.}(2003)\citenamefont
  {Chevillard}, \citenamefont {Roux}, \citenamefont {L{\'e}v{\^e}que},
  \citenamefont {Mordant}, \citenamefont {Pinton},\ and\ \citenamefont
  {Arn{\'e}odo}}]{chevillard2003}%
  \BibitemOpen
  \bibfield  {author} {\bibinfo {author} {\bibfnamefont {L.}~\bibnamefont
  {Chevillard}}, \bibinfo {author} {\bibfnamefont {S.~G.}\ \bibnamefont
  {Roux}}, \bibinfo {author} {\bibfnamefont {E.}~\bibnamefont
  {L{\'e}v{\^e}que}}, \bibinfo {author} {\bibfnamefont {N.}~\bibnamefont
  {Mordant}}, \bibinfo {author} {\bibfnamefont {J.-F.}\ \bibnamefont
  {Pinton}},\ and\ \bibinfo {author} {\bibfnamefont {A.}~\bibnamefont
  {Arn{\'e}odo}},\ }\bibfield  {title} {\bibinfo {title} {{L}agrangian velocity
  statistics in turbulent flows: Effects of dissipation},\ }\href@noop {}
  {\bibfield  {journal} {\bibinfo  {journal} {Phys. Rev. Lett.}\ }\textbf
  {\bibinfo {volume} {91}},\ \bibinfo {pages} {214502} (\bibinfo {year}
  {2003})}\BibitemShut {NoStop}%
\bibitem [{\citenamefont {Corbetta}\ \emph {et~al.}(2021)\citenamefont
  {Corbetta}, \citenamefont {Menkovski}, \citenamefont {Benzi},\ and\
  \citenamefont {Toschi}}]{corbetta2021deep}%
  \BibitemOpen
  \bibfield  {author} {\bibinfo {author} {\bibfnamefont {A.}~\bibnamefont
  {Corbetta}}, \bibinfo {author} {\bibfnamefont {V.}~\bibnamefont {Menkovski}},
  \bibinfo {author} {\bibfnamefont {R.}~\bibnamefont {Benzi}},\ and\ \bibinfo
  {author} {\bibfnamefont {F.}~\bibnamefont {Toschi}},\ }\bibfield  {title}
  {\bibinfo {title} {Deep learning velocity signals allow quantifying
  turbulence intensity},\ }\href@noop {} {\bibfield  {journal} {\bibinfo
  {journal} {Science Adv.}\ }\textbf {\bibinfo {volume} {7}},\ \bibinfo {pages}
  {eaba7281} (\bibinfo {year} {2021})}\BibitemShut {NoStop}%
\bibitem [{\citenamefont {Dombre}\ and\ \citenamefont
  {Gilson}(1998)}]{dombre1998}%
  \BibitemOpen
  \bibfield  {author} {\bibinfo {author} {\bibfnamefont {T.}~\bibnamefont
  {Dombre}}\ and\ \bibinfo {author} {\bibfnamefont {J.-L.}\ \bibnamefont
  {Gilson}},\ }\bibfield  {title} {\bibinfo {title} {Intermittency, chaos and
  singular fluctuations in the mixed {Obukhov}-{Novikov} shell model of
  turbulence},\ }\href
  {https://doi.org/https://doi.org/10.1016/S0167-2789(97)80015-2} {\bibfield
  {journal} {\bibinfo  {journal} {Physica D: Nonlinear Phenomena}\ }\textbf
  {\bibinfo {volume} {111}},\ \bibinfo {pages} {265} (\bibinfo {year}
  {1998})}\BibitemShut {NoStop}%
\bibitem [{\citenamefont {Mailybaev}(2013)}]{mailybaev2013}%
  \BibitemOpen
  \bibfield  {author} {\bibinfo {author} {\bibfnamefont {A.~A.}\ \bibnamefont
  {Mailybaev}},\ }\bibfield  {title} {\bibinfo {title} {Blowup as a driving
  mechanism of turbulence in shell models},\ }\href
  {https://doi.org/10.1103/PhysRevE.87.053011} {\bibfield  {journal} {\bibinfo
  {journal} {Phys. Rev. E}\ }\textbf {\bibinfo {volume} {87}},\ \bibinfo
  {pages} {053011} (\bibinfo {year} {2013})}\BibitemShut {NoStop}%
\end{thebibliography}

%

\end{document}